\documentclass[twocolumn,superscriptaddress,amsfonts]{revtex4-1}
\usepackage{epsfig}
\begin{document}
\title{Spectra in nested Mach-Zehnder interferometer experiments}
\author{Marcin Wie\'sniak}\affiliation{Institute of Informatics, Faculty of Mathematics, Physics, and Informatics,\\ University of Gda\'nsk, 80-308 Gda\'nsk, Poland}
\begin{abstract}
By the means of the standard quantum mechanics formalism I present an explicit derivation of the structure of  power spectra in Danan {\em et al.} and Zhou {\em et al.} experiments with nested dynamically changing Mach-Zehnder interferometers. The analysis confirms that we observe prominent, first-order peaks on frequencies related to some of the elements of the interferometer, but not on others. However, as I shall demonstrate, there are also other, weaker effects related to all relevant elements of the setup. In case of the Danan {\em et al.} setup, there are even peaks at all frequencies of element oscillations. When confronted in an experiment, these observations shall challenge the interpretation of the experiments based on anomalous trajectories of light.
\end{abstract}
\maketitle
Interference of light is one of its most intriguing features, with a number applications in various areas. It reveals the wave nature of electromagnetic radiation, which coexists with particle-specific effects. Interference allows to recombine a spatially split signal (be it, a wave or individual particles) in such a way, that it can be deterministically redirected to a specific output. However, any attempt to learn a specific path taken by the signal within the interferometer immediately causes a decrease in the interference pattern depth due to the Welcher-Weg information complementarity \cite{COMPL}.

More recently, an experiment was conducted, supposedly giving some new insight to the problem of the interference. 
Vaidman \cite{VAIDMAN0} has suggested an experimental setup, in which one Mach-Zehnder interferometer is nested in one of the arms of another. The setup was subsequently realized in Ref. \cite{VAIDMAN} and more recently, the experiment was redone in a slightly modified scheme \cite {F08}, but confirming results of Danan, Farfunik, Bar-Ad, and Vaidman, also another group \cite{Alonso} conducted a similar experiment. These results were actively discussed in recent years (Ref. \cite{VAIDMAN} was chosen for {\em Physical Review Letters Viewpoint} at the time of its publication and by June, the 14th, 2018, it	 was cited 63 times, according to Web of Science\textsuperscript{\textregistered}. The most noticeable critical comments include Refs. \cite{C1,C2,C3,C4,C5,C6,C7,C8}). A very similar concept also appears in the  context of ``counterfactual communication'' \cite{CF1,CF2}.

The details of these two experiments are described below and depicted in FIGS. 1 and 2. In the version from Ref. \cite{VAIDMAN}, we have a large Mach-Zehnder interferometric loop. The source of coherent light is at the top in FIG. 1 and the light falls onto the first beam splitter (BS) with transmitivity $1/3$. If it is transmitted, it then falls on mirror C and on the second BS, also of transmitivity $1/3$. At one of the outputs of this BS there is a quad-cell detector, allowing to register the difference of intensities of light falling on its upper and lower parts (with the respect to the plane of the interferometer).  If light reflects from the first BS, it is first redirected downwards (in terms of the figure) by mirror E. Then it enters a smaller Mach-Zehnder loop, with two balanced BSs and mirrors A and B at the corners \cite{FN1}. Should the light leave this loop horizontally, it would encounter mirror F that redirects it onto the second BS to recombine with the initially transmitted beam. However, initially, the smaller loop is set to redirect all the light incoming from the direction of mirror F to the other output,  where it is dumped.

The mirrors in this configuration oscillate with respective frequencies $f_A,...,f_F$ by being slightly tilted with respect to axes in the plane of both interferometers, causing the beam to be displaced somewhat differently along each path. The experimentalists study the power spectrum of the aforementioned difference in intensities. They observe peaks at frequencies $f_A, f_B$, and $f_C$, which are obvious certificates of interaction of photons with these mirrors, but not at $f_E$ or $f_F$ \cite{FN2}. 
\begin{figure}[htbp]
	\centering
		\includegraphics[width=0.45\textwidth]{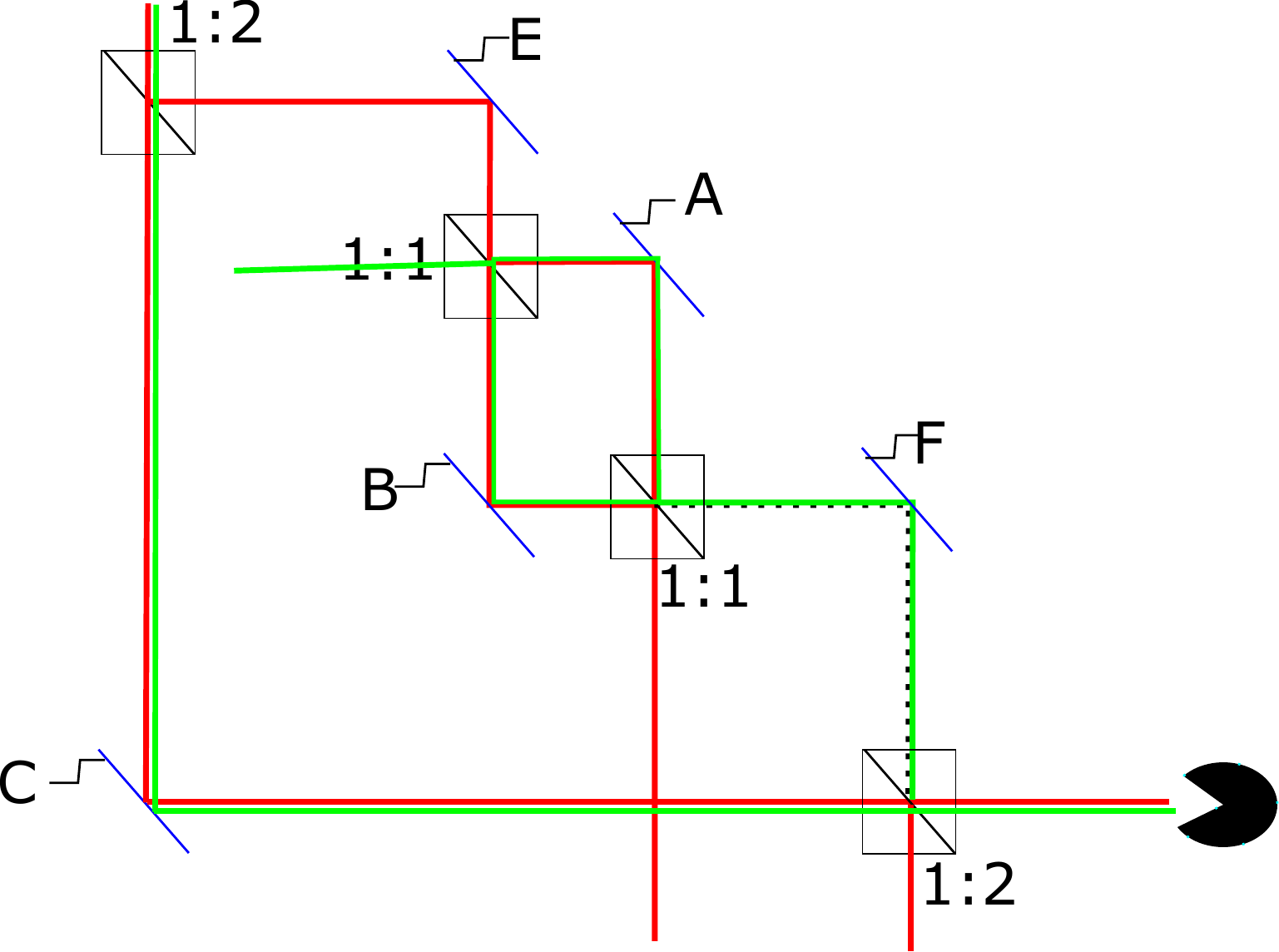}
	\caption{The experimental setup in Ref. \cite{VAIDMAN}. Mirrors A, B, C, E, F are periodically tilted with frequencies $f_A,...,f_F$, causing a small displacement of the beam in direction perpendicular to the plane of figure. The red and the green line would correspond to the forward- and backward-evolving state in the two-state vector formalism \cite{TSVF1,TSVF2}.}
	\label{FIG0a}
\end{figure}
 
In the setup of Ref. \cite{F08}, all the mirrors are stationary. There are no mirrors E or F, the first beamsplitter redirects a portion of light directly to the smaller loop, and one of its outputs points directly towards the recombination on the second beam-splitter. The small loop is, again, tuned destructively, and electro-optical modulators (EOMs) are placed on each path between each pair of BSs. They modulate the phase with very small amplitudes and frequencies $f_A,...,f_F$. This test is meant for certification of presence of photons along paths, rather than at mirrors, but again, we observe a lack of certain peaks in the power spectrum, and similar conclusions are drawn. 
\begin{figure}
	\centering
		\includegraphics[width=0.45\textwidth]{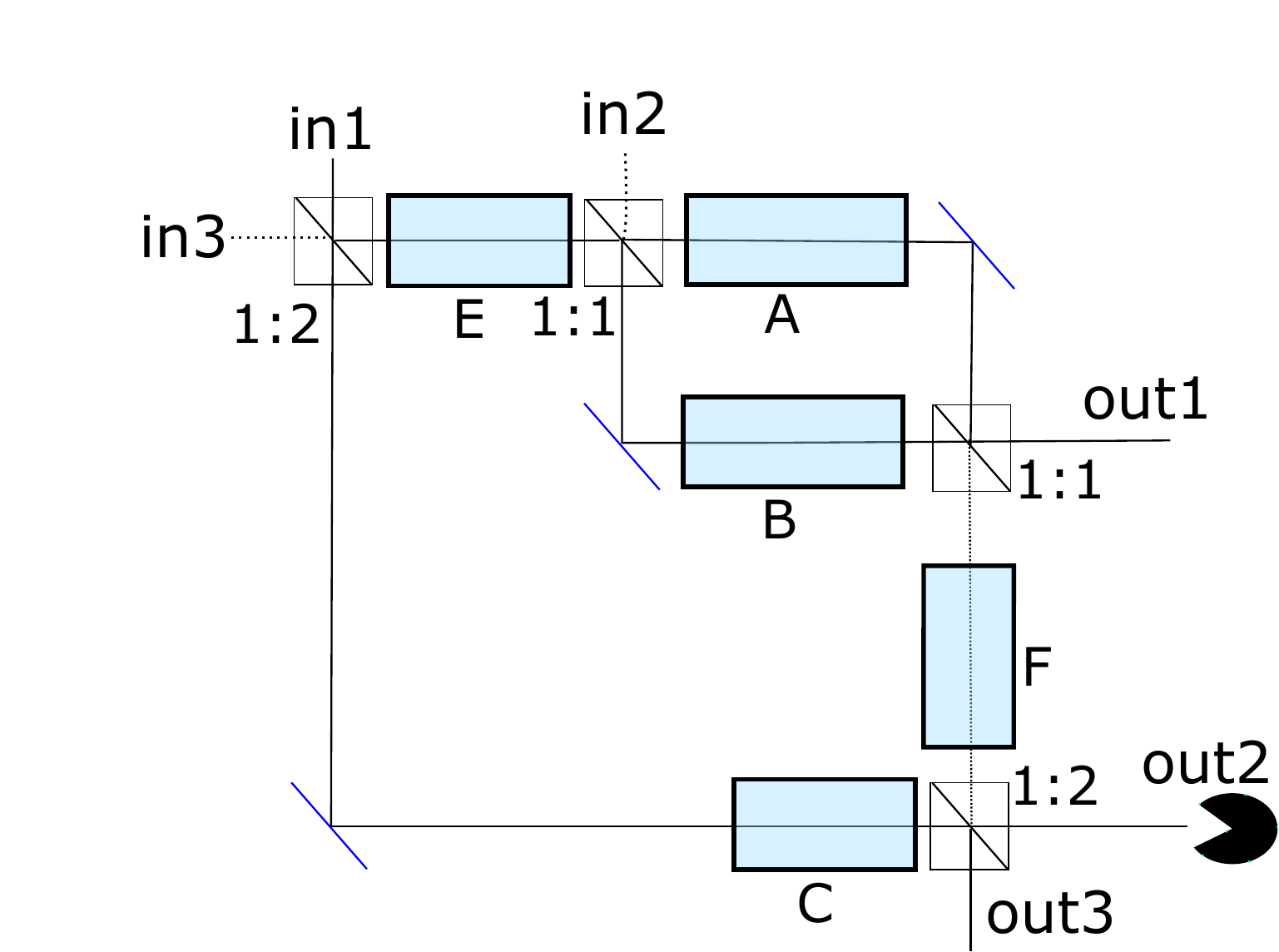}
	\caption{The experimental setup from Ref. \cite{F08}. Electro-optical modulators modulate phases of the beams with frequencies $f_A,...,f_F$. Input and output channels are $in1, in2, in3, out1, out2, out3$} 
	\label{fig:FIG0b}
\end{figure}

In this letter I present a simple quantum-mechanical analysis of these experiments. It will be demonstrated below that, indeed, in both experiments, first-order peaks at frequencies $f_E$ and $f_F$ are absent, but there are other certificates of presence of light in the whole setup, namely lower-order peaks on frequencies being combinations of all frequencies involved. While these peaks may be significantly lower, they certainly prevent us from formulating strong claims about the absence of light at certain locations.

\begin{figure}
	\centering
		\includegraphics[width=0.45\textwidth]{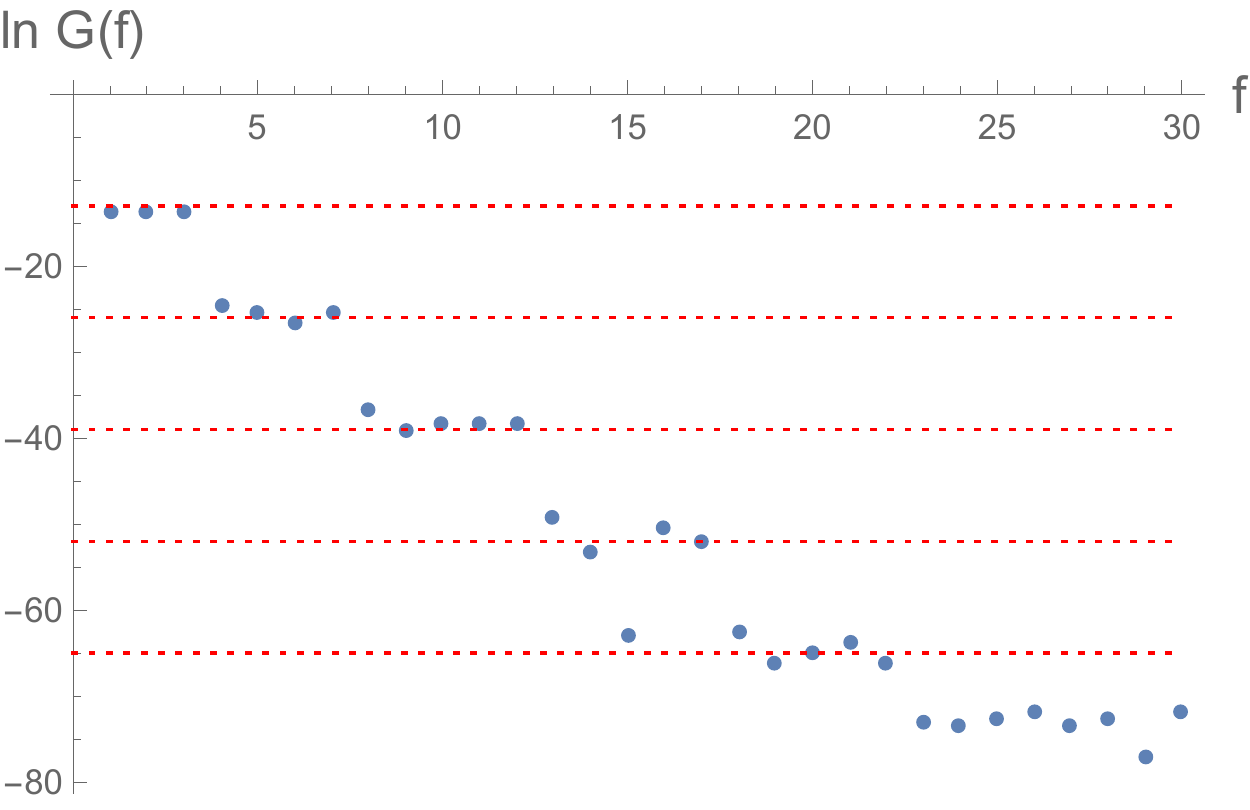}
	\caption{The logarithm of the normalized discrete power spectrum $G(f)$ of the signal reaching the detector in the Zhou {\em et al.} experiment. Equidistant red dotted lines serve as eye guides and indicate the order of the peak (up to the numerical precision). Notice the second-order peaks in range $4=f_F-f_A=f_E-f_A$ to $7=f_F+f_B$. The assumed values are $A_0=\pi/500, f_A=1, f_B=2, f_C=3, f_E=4, f_F=5$.}
	\label{FIG0c}
\end{figure}

I start with analyzing the scheme of Ref. \cite{F08}, which is more straight-forward. Between beam-splitters we place EOMs, which simply introduce phase factors. I assume light entering the setup through the first input is either a heralded single photon or a simply a bright coherent state. I shall also assume temporarily assume perfect detectors without dark counts. The transformation between input and output mode annihilation operators realized by the interferometer is
\begin{eqnarray}
\label{interf}
\left(\begin{array}{c}\hat{a}^{\text{out}}_1\\\hat{a}^{\text{out}}_2\\\hat{a}^{\text{out}}_3\end{array}\right)&\leftarrow&\left(\begin{array}{ccc}1&0&0\\0&\sqrt{2/3}&\sqrt{1/3}\\0&\sqrt{1/3}&-\sqrt{2/3}\end{array}\right)\left(\begin{array}{ccc}1&0&0\\0&e^{i\phi_F(t)}&0\\0&0&1\end{array}\right)\nonumber\\
&\times&\left(\begin{array}{ccc}\sqrt{1/2}&\sqrt{1/2}&0\\\sqrt{1/2}&-\sqrt{1/2}&0\\0&0&1\end{array}\right)\left(\begin{array}{ccc}e^{i\phi_A(t)}&0&0\\0&e^{i\phi_B(t)}&0\\0&0&1\end{array}\right)\nonumber\\
&\times&\left(\begin{array}{ccc}\sqrt{1/2}&1/\sqrt{2}&0\\\sqrt{1/2}&-\sqrt{1/2}&0\\0&0&1\end{array}\right)\left(\begin{array}{ccc}e^{i\phi_E(t)}&0&0\\0&1&0\\0&0&e^{i\phi_C(t)}\end{array}\right)\nonumber\\
&\times&\left(\begin{array}{ccc}\sqrt{2/3}&0&\sqrt{1/3}\\0&1&0\\\sqrt{1/3}&0&-\sqrt{2/3}\end{array}\right)\left(\begin{array}{c}\hat{a}^{\text{in}}_1\\\hat{a}^{\text{in}}_2\\\hat{a}^{\text{in}}_3\end{array}\right),
\end{eqnarray}
where $\phi_X=A_0\sin(2\pi f_X t)$, $X=A,B,C,E,F$, and $A_0$ is the common amplitude of the phase change. The signal is fed by the first input. 
Consider the probability amplitude that a photon entering the setup reaches the detector, or the amplitude of the coherent light at the second output.
\begin{eqnarray}\label{Ft}
V_2(t)&=&\frac{1}{3}\left(e^{iA_0\sin(2\pi f_Ct)}\right.\nonumber\\
&+&e^{iA_0(\sin(2\pi f_At)+\sin(2\pi f_Et)+\sin(2\pi f_Ft))}\nonumber\\
&-&\left.e^{iA_0(\sin(2\pi f_Bt)+\sin(2\pi f_Et)+\sin(2\pi f_Ft))}\right),
\end{eqnarray}
where $A_0$ is the common amplitude of phase change.
The result of the experiment analyzed in Ref. \cite{F08} is the power spectrum of this signal, which is the decomposition of it into oscillating terms over a finite duration. For convenience, I choose all frequencies to be multiples of a certain base frequency, so that it is possible to make a decomposition into discrete frequencies. This is the (renormalized) discrete energy spectrum over a single period, which would be the power spectrum in the limit of the experiment running infinitely long. In a real-life experiment, it would mean that the peaks are very narrow and well-defined. The figure of merit would hence be 
\begin{equation}
G(f)=\left|\int_0^1 e^{-2\pi i f t}V_2(t)dt\right|^2\quad(f=0,1,2,...).
\end{equation}
This integral can be treated analytically in the following fashion. First, we use the Maclaurin expansion of the exponent, $\exp(x)=\sum_{k=0}^\infty\frac{x^k}{k!}$. Then we iteratively use identities $\sin(x)\sin(y)=\frac{1}{2}[-\cos(x+y)+\cos(x-y)]$, $\cos(x)\cos(y)=\frac{1}{2}[\cos(x+y)+\cos(x-y)]$, and $\sin(x)\cos(y)=\frac{1}{2}[\cos(x+y)+sin(x-y)]$ we see that apart from a peak of magnitude $n$ (an arbitrary constant) at 0, $G(f)$ will have peaks of magnitude $\frac{1}{3}nA_0^2$ at $f_A$, $f_B$ and $f_C$, as well as peaks of magnitude $\frac{1}{12}nA_0^4$ at $2f_A$, $2f_B$, $2f_C$, and $f_{A/B}\pm f_{E/F}$. In general, peaks will be present at frequencies $n_1f_C$, $n_2 f_A+n_3 f_E+n_4 f_F$, and $n_2 f_B+n_3 f_E+n_4 f_F$, where $n_1$, $n_2$, $n_3\neq 0$, and $n_4\neq 0$ are integers. An example of $G(f)$ is shown in FIG. \ref{FIG0c}.
If we take into account the dark counts of the detectors, we shall assume that they happen spontaneously, in accordance to the Poissonian distribution. This would overlay the spectrum onto the spectrum of the (pink) noise, which is characterized by the exponential decay for high frequencies. Nevertheless, the frequencies discussed here should still be distinguished for amounts of dark counts attainable with modern detectors.

\begin{figure}
	\centering
		\includegraphics[width=0.45\textwidth]{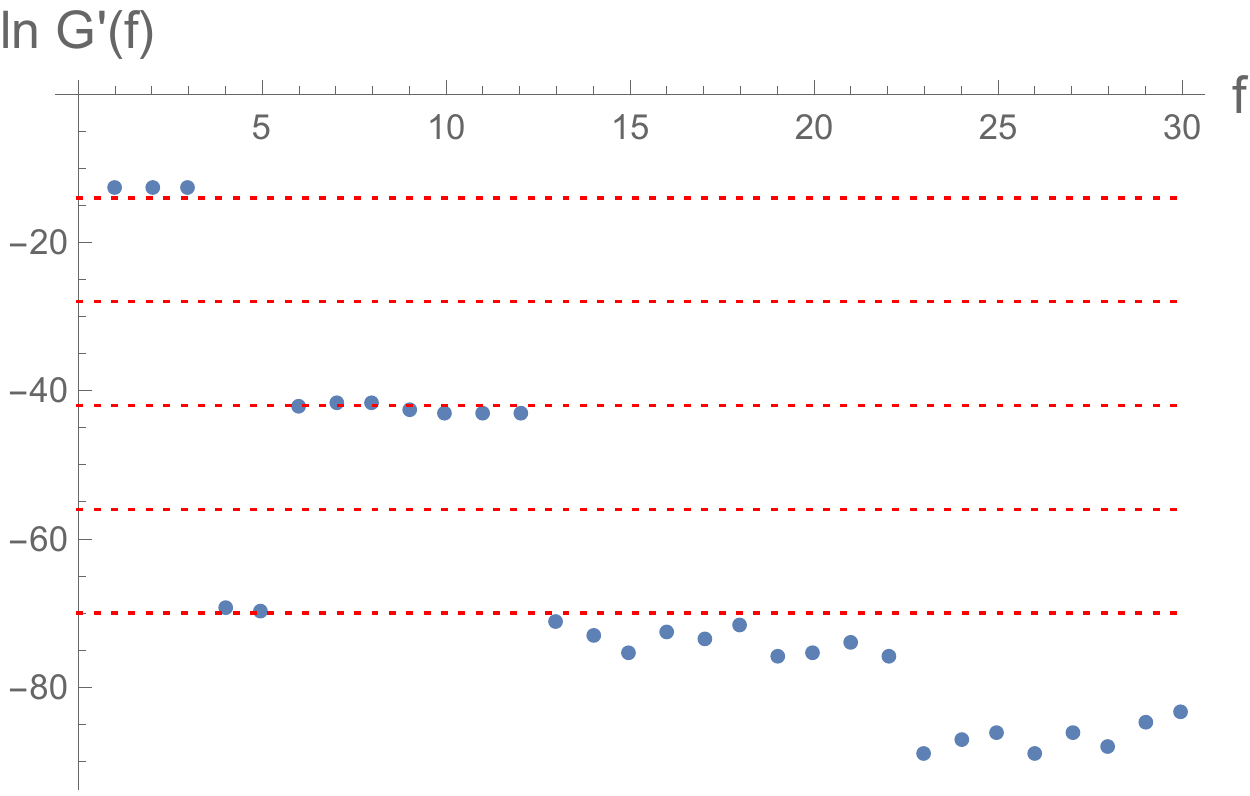}
	\caption{The logarithm of the normalized discrete power spectrum $G'(f)$ of the signal difference $I(t)$ in the Danan {\em et al.} experiment, as given by Eq. (\ref{IofT}). Equidistant red dotted lines serve as eye guides and indicate the order of the peak (up to the numerical precision). In the first order we peaks at $f_A, f_B$, and $f_C$. Peaks at $f_E$ and $f_F$ are of the fifth order, other ($f_E$- and $f_F$-related) frequencies have third-order peaks. The assumed values are $\sigma/\delta=1000, f_A=1, f_B=2, f_C=3, f_E=4, f_F=5$.}
	\label{FIG0d}
\end{figure}

I now pass to deriving a similar result for the Danan-Farfunik-Bar-Ad-Vaidman experiment \cite{VAIDMAN}. Therein, the mirrors establishing the corners of the interferometer were slightly tilted, effectively causing a small displacement of the beam  in direction $d$, perpendicular to the plane of the experiment. Let me assume that a displacement caused by each mirror is of a common magnitude, $\delta$, but the result could easily be rewritten for arbitrary shifts \cite{FN3}. For simplicity, I shall also assume the Gaussian profile of the incoming beam, $\psi(x,y)\propto\exp{-\frac{x^2+y^2}{2\sigma^2}}$. The Hilbert space is now a product of a three-dimensional space describing in which path of the interferometer light is found, and the continuous space of the transversal modes. The phase shift factors are now replaced by displacement operators, $D_X(t)=\exp\left[-\delta\sin(2\pi f_Xt)\frac{\partial}{\partial y}\right]$. Plugging these changes into Eq. (\ref{interf}) we get (up to reordering the of the output modes)
\begin{equation}
\label{v2prim}
V'_2(t)\nonumber=\frac{1}{3}\left(D_{C}(t)+D_{E,F,A}(t)-D_{E,F,B}(t)\right)\psi(x,y)
\end{equation}
with $D_{E,F,A}(t)=D_F(t)D_A(t)D_E(t)$ and likewise for $D_{E,F,B}$. In Ref. \cite{VAIDMAN}, the quantity studied was the difference between the intensities of the upper and the lower part of the beams.
\begin{equation}
\label{Iofy}
I(t)=\int_{-\infty}^{\infty}dx\int_{-\infty}^{\infty}dy\text{sign}(y)|V'_2(t)|^2,
\end{equation}
which, after plugging in Eq. (\ref{v2prim}), reads
\begin{eqnarray}
\label{IofT}
&&I(t)\nonumber\\
\propto&&\text{Erf}\left(\frac{d_1}{\sigma}\right)+\text{Erf}\left(\frac{d_2}{\sigma}\right)+\text{Erf}\left(\frac{d_1}{\sigma}\right)\nonumber\\
+&&e^{-\frac{(d_1-d_2)^2}{4\sigma^2}}\text{Erf}\left(\frac{d_1+d_2}{2\sigma}\right)+e^{-\frac{(d_1-d_3)^2}{4\sigma^2}}\text{Erf}\left(\frac{d_1+d_3}{2\sigma}\right)\nonumber\\
+&&e^{-\frac{(d_2-d_3)^2}{4\sigma^2}}\text{Erf}\left(\frac{d_2+d_3}{2	\sigma}\right),
\end{eqnarray}
where a shorthand notation was introduced, $d_1=\delta\sin(2\pi f_C t)$, $d_2=\delta[\sin(2\pi f_A t)+\sin(2\pi f_E t)+\sin(2\pi f_F t)]$, and $d_3=\delta[\sin(2\pi f_B t)+\sin(2\pi f_E t)+\sin(2\pi f_F t)]$. Let us now refer to the Maclaurin expansion of the function,
\begin{equation}
I(t)\propto d_1+d_2-d_3-\frac{2}{\sigma^2}(d_1^3+d_2^3-d_3^3)+...\,.
\end{equation}
Using this expansion we see that in the first order of $\delta$ the spectrum $G'(f)$ of $I(t)$, where
\begin{equation}
G'(f)=\left|\int_0^1I(t)\sin(2\pi f t)dt\right|^2,
\end{equation}
will still have peaks at $f_A$, $f_B$, and $f_C$, but not at $f_E$ and $f_F$. In the third order, we will have peaks at $\pm f_Y$, $\pm 3f_Y$ ($Y=A,B,C$), $\pm f_{A/B}\pm 2 f_E/F$, $\pm 2f_{A/B}\pm f_{E/F}$, and $\pm f_{A/B} \pm f_E\pm f_F$, where, e.g., $f_{A/B}$ states for an arbitrary alternative between $f_A$ and $f_B$, etc.. Interestingly, we  have $\frac{\partial^5I(t)}{\partial d_2^5}\neq -\frac{\partial^5I(t)}{\partial d_3^5}$, so we could observe direct fifth-order peaks at $f_E$ and $f_F$. An example of a spectrum in this experiment is depicted in FIG. \ref{FIG0d}.

It shall be noted that the Authors of Refs. \cite{VAIDMAN} and \cite{F08}, while they acknowledge (and disregard) even the classical description of the result,	 present their own formalism explaining this phenomenon, known as two state vector formalism (TSVF) \cite{TSVF1,TSVF2}. It requires a consideration of two states: preselected by our knowledge of the initial state (the source), which is forward-evolved in time, and postselected, defined by the outcome of the measurement, which would be back-evolved. In this formalism, we overlap these two states and {\em observable} (first-order) effects appear only at these locations, which a propagating particle could reach {\em propagating both forward and backward}. In Ref. \cite{VAIDMAN} an interpretation that a particle was present {\bf only} at these locations was assumed. A more relaxed version was proposed in Ref. \cite{F12}, where Vaidman suggests that while photons manifested their presence at inside the smaller interferometer or at mirror C, they were only {\em secondarily present} at and near mirrors E and F (or respective EOMs), since even though we do not observe peaks at $f_E$ or $f_F$ in the spectrum, attenuators placed there can alter heights of the peaks at $f_A$ and $f_B$. 

It was seemingly the the TVSF that help the Authors of Refs. \cite{VAIDMAN,F08} to understand their results and encouraged their speculations. This formalisms is still seen handful for interpreting experiments described in influential journals \cite{Rev1,Rev2,Rev3}.

I shall also present the analysis of the variant of the experiment, in which the lowest path is blocked. In the realization of Danan {\em et al.}, the smaller MZI has a leakage due to an uneven displacement of the beams. The difference of intensities reaching the upper and the lower halves of the detector is proportional to 
\begin{eqnarray}
I'(t)\propto&&\pi\sigma^2\left(\text{Erf}\left(\frac{d_2}{\sigma}\right)+\text{Erf}\left(\frac{d_3}{\sigma}\right)\right.\nonumber\\
-&&\left.2e^{-\frac{(d_2-d_3)^2}{4\sigma}}\text{Erf}\left(\frac{d_2+d_3}{2\sigma}\right)\right)\nonumber\\
=&&\frac{\sqrt{\pi}}{12\sigma^3}\left((d_3^5+d_2^5)+(d_2^4d_3+d_2d_3^4)-2(d_2^3d_3^2+d_2^2d_3^3)\right)\nonumber\\
+&&...\, .
\end{eqnarray} 
Thus we again see peaks of the fifth order at frequencies which may involve all of $f_A, f_B, f_E$, and $f_F$, and in particular, at each of these frequencies. Note that some of the frequencies are combinations of $f_A$ and $f_B$.

In the other experiment, Zhou {\em et al.} see peaks at $f_A$ and $f_B$. These two frequencies modulate the detuning of the smaller MZI, while $f_E$ and $f_F$ now together modulate the global phase of the beam. Registering these frequencies in the optical domain would require interference with a reference beam, which contradicts the lower arm being blocked.  

The modification in Ref. \cite{Alonso} of the scheme of Ref. \cite{VAIDMAN} was to put dove prisms in the arms of the smaller MZI. By aligning the prisms, Alonso and Jordan were able to see a (first-order) peak at $f_E$, but not at $f_F$. The Authors correctly identify that the dove prism contribute to detuning of the smaller MZI. The role of the dove prisms is to mirror-reflect a beam on path A with respect to beam on path B. Then, the shift caused by mirror E becomes the opposite on paths A an B. The signal difference can be reproduced by plugging $d_3=\delta[-\sin(2\pi f_E t)-\sin(2\pi f_B t)+\sin(2\pi f_F t)]$ to Eq. (\ref{IofT}). $f_E$-dependent terms in the Maclaurin expansion do not cancel, hence the peak at $f_E$. As the detuning is amplified, more light can be redirected towards mirror F, but still, terms dependent solely on $f_F$ will cancel in the first and the third order.

Finally, let me remark that the same analysis will hold to any periodic modulations of phases or displacements; one would simply need to employ the formalism of the Fourier sums.
	
In conclusion, I have shown that, indeed, in experiments on nested Mach-Zehnder interferometers, described in Refs. \cite{VAIDMAN} and \cite{F08}, first-order peaks at $f_E$ and $f_F$ are missing in power/energy spectra. However, a straight-forward interferometric calculation show that this lack is a simple effect of destructive interference, as first noted in Ref. \cite{C4}. Other effects related to these frequencies are present, however. Moreover, in the case of the Danan-Farfunik-Bar-Ad-Vaidman setup, we can directly observe fifth-order peaks at the frequencies of interest. With a choice of a small phase shift or beam displacement, the peaks related to $f_E$ or $f_F$ are very weak, being of the second and the third order of the modulation amplitude. However, with the state-of-the art-technology they are still feasible to be observed, and their anticipation should help us understand the laws of light propagation better, as well as prevent us from formulating definite statements about absence of light at certain locations. Also, introducing the notion of	 {\em secondary presence} does not seem to contribute to explaining these effects within TSVF, and Electrodynamics does not give any foundations of such a gradation. 

The effects predicted here now await to be verified in future nested MZI experiments. Once they are confirmed, Two-vector state formalism must be redeveloped to describe them (or otherwise it will be unable to deliver the full description of an experiment). This, in turn, will allow to reexamine the definitions of ``past'' and ``presence'' (understood as an event of a particle being then and there), which shall take into accounts all events.

{\em (quant-ph only:)}Studying ultra-weak or higher-order effects has paved us ways to confirm phenomena such as gravitational lensing \cite{LENSING}, 
creation of photon pairs through parametric down-conversion \cite{WEINBERG}, exoplanets \cite{EXO}, solar neutrino detection \cite{NEUT}, general relativistic redshift \cite{TIMESHIFT}, or gravitational waves \cite{GWAVES}. Curiously, now they can be used to verify the most basic laws of light propagation. 

This work was supported by the Polish National Center for Science (NCN) Grant No. UMO-2015/19/B/ST2/01999 (task 1).


\begin{thebibliography}{99}
\bibitem{FN1} Ocassionally, labels A, B, C will be also used for paths.
\bibitem{FN2} At this point, the Reader is kindly invited to examine conclusions of Refs. \cite{VAIDMAN,F08}.
\end{thebibliography}
\end{document}